\documentclass[floats,epsfig,pdflatex,twocolumn]{revtex4}
\usepackage{graphics,epsfig,graphicx}
\usepackage{color}
\usepackage{makeidx}
\usepackage{latexsym}
\usepackage{amsmath}
\usepackage{array}
\usepackage{subfigure}

\def\bra#1{\mathinner{\langle{#1}|}}
\def\ket#1{\mathinner{|{#1}\rangle}}
\newcommand{\braket}[2]{\langle #1|#2\rangle}

\begin{document}

\title{Signature of Non-Abelian to Abelian Transition in Spin Systems Through Geometric Phase}

\author{D. De Munshi\footnote{debashis.de.munshi@nus.edu.sg} and Manas Mukherjee\footnote{phymukhe@nus.edu.sg}}

\affiliation{Centre for Quantum Technologies, National University of
Singapore, Singapore - 117543}

\begin{abstract}
Abelian and Non-Abelian evolution of a quantum system manifests
differently in the geometric phase acquired by the system under such
evolutions. In this work we develop and study, using dressed state
techniques, an experimentally realizable spin system which allows us
to transit smoothly from non-Abelian to Abelian evolutions by
changing externally controllable parameters. The study provides
insights into the underlying physical phenomenon governing such a
transition allowing us greater control on phase generation in a
quantum system. The robustness of geometric phase against
fluctuations of the external parameters of the system has also been
studied in this work. The noise analysis has direct consequence on
the present search for fault tolerant quantum gates using geometric
phase.

\end{abstract}
\maketitle

\section{Introduction}

Geometric phase originates due to cyclic time evolution of an
Hamiltonian. This phase can be distinguished from the usual
dynamical phase by its dependence on the quantum level structure of
the system and the form of the time evolution of the involved
Hamiltonian. Although geometric phase depends on the quantum level
structure, it is independent of the energy eigenvalues. Therefore it
is largely conceived as immune to external perturbations and hence a
good quantum computation resource
\cite{Zanardi99},\cite{Ekert00},\cite{Duan01},\cite{Shi03},\cite{Cosmo09}.
Any quantum computation at its basic level requires quantum gates
for both single and two qubits. In order to perform reliably as a
component of a quantum computation system, these quantum gates needs
to have an error generation probability of $10^{-4}$, which is known
as fault tolerant threshold (FTT)\cite{Knill10}. So far the only
single qubit operation that has shown such kind of FTT is based on
an ion trap experiment \cite{Brown11} . Lemmer {\it et. al}
\cite{Lemmer13} proposed a protocol based on spin phonon interaction
which has the potential to beat the FTT. However, their analysis is
specific to the proposed qubit system. Geometric phase, being
inherently robust to external field fluctuations, provide a likely
resource which can be exploited to create quantum gates having such
FTT.\\

In general, the robustness of geometric phase is an inherent
property of the mechanics itself. Therefore understanding the origin
of such robustness in the context of geometric phase in spin system
can allow more controllability on such geometric phase gates. Any
universal quantum computation will need single and two qubit gates
which are both Abelian and non-Abelian in nature\cite{Zanardi99}. In
most cases, studies on geometric phases are restricted to either of
these two regimes
\cite{Jacob07},\cite{Vaya11},\cite{Wang12},\cite{Jonas09}. Here we
put forward a study in which we can go from Abelian to non-Abelian
regime in the same spin system by slowly breaking the symmetry of
the system. Analysis of such a system not only provides insights
into the underlying physical processes governing each of the limits
but also effectively probes the limits of the adiabatic theorem
\cite{Marz04},\cite{Bould08} and the relation of adiabaticity and
non-abelian behaviour. This is relevant as most of the quantum
states are operated in the super-adiabatic regime \cite{Zu14}. Our
analysis also provides room for identifying the relevant parameters
which influence the phase fluctuation in different regimes.

\section{Theory}


In \cite {Berry84}, Berry formulated the form of geometric phase
under adiabatic approximation as

\begin{equation}
\mathcal{G}_{n}=i\oint\bra{\psi_{n}}\nabla_{r}\ket{\psi_{n}}dR
\end{equation}
for the $n$th eigenstate. The quantity
$\gamma_{n}=\bra{\psi_{n}}\nabla_{r}\ket{\psi_{n}}$ is called the
gauge of the evolution because it remains invariant under any
similarity transformation, except those involving the variable of
the evolution themselves. This definition of the 'scalar' gauge
holds only for non-degenerate levels. For degenerate levels, the
definition is generalized to a matrix gauge
$\gamma_{mn}=i\bra{\psi_{m}}\nabla_{r}\ket{\psi_{n}}$, where $m$ and
$n$
belong to the degenerate subspace \cite{Wilc84}.\\


The adiabatic form of geometric phase can also be derived from the
adiabatic theorem. The probability amplitude of an eigenstate
belonging to a time dependent Hamiltonian varies as

\begin{equation}
\dot{C_{m}}=-C_{m}\braket{\psi_{m}}{\dot{\psi_{m}}}-\sum_{n\neq m}
C_{n}\frac{\bra{\psi_{m}}\dot{H}\ket{\psi_{n}}}{E_{n}-E_{m}}~
e^{i(\xi_{n}-\xi_{m})},
\label{basic_eq}
\end{equation}

where the states $m$ and $n$ are non-degenerate and $\xi_{m}$ and
$\xi_{n}$ are the dynamical phases. The above equation governs the
time dependence of the amplitudes of the states, beyond the
dynamical contribution. Under adiabatic approximation,
$\bra{\psi_{m}}\dot{H}\ket{\psi_{n}} \ll (E_{n}-E_{m})$ and hence
the second term in Eq. \ref{basic_eq} can be neglected in comparison
to the first term. Thus for adiabatic evolution, there is no
'mixing' of the different eigenstates. However, as a consequence of
the time dependence of the Hamiltonian, there is an additional
phase, on top of the dynamical phase, governed by
$\braket{\psi}{\dot{\psi}}$. Under the conditions of implicit time
dependence, this term leads to the underlying gauge of geometric
phase as derived by Berry. In the following sections we expand the
ideas of Abelian and non-Abelian evolution and
their corresponding gauges.\\

{\bf Abelian:} Abelian evolution corresponds to evolution without
any population transfer. For a non-degenrate set of levels, under
adiabatic condition, all evolutions are Abelian, since adiabaticity
guarantees lack of population transfer or mixings. Degenerate levels
can also have Abelian evolutions, if the underlying gauge matrix
corresponding to the degenerate subspace is diagonal.\\

{\bf Non-Abelian:} On the contrary, non-Abelian evolutions
inherently introduces mixing of states or population transfer
between the states. For a degenerate subspace, if the off-diagonal
elements of the gauge matrix are nonzero, then the evolution is
considered as non-Abelian \cite{Wilc84}. However, degeneracy itself
doesn't guarantee non-zero off diagonal elements. A subspace of
non-degenerate levels, under certain evolutions can have non-zero
off diagonal elements. However, conditions imposed by adiabaticity
doesn't leave room for population transfer in such cases and the
off-diagonal elements have no physical significance under such
conditions. This can be easily demonstrated from Eq. \ref{basic_eq}
as the coupling term drops off because of finite strength of the
oscillatory function and the relatively smaller coupling strength in
the adiabatic limit.\\


{\bf Non-Abelian to Abelian Transition:} The primary goal of this
work is to see how a system responds if it is taken in a continuous
manner from non-Abelian to Abelian evolution and vice versa. To
achieve this, we need an evolution with degeneracies and focus on
non-Abelian degenerate subspaces. Now, if we can introduce
non-degeneracy into the system, without changing the underlying
geometry and hence the gauge matrix of the evolution, then we can
observe the physical significance of the off-diagonal elements
slowly diminishing and vanishing in the Abelian regime. Thus the
primary goal is to study the dynamics of the off-diagonal elements
with respect to the symmetry breaking field. This also allows us to
probe the Adiabatic theorem the limits of which, under the influence
of a symmetry breaking field has long been
debated\cite{Marz04},\cite{Bould08}\\.

\section{System}


The system we work with is a spin system interacting with
electro-magnetic fields. Even though we are interested in the
$D_{3/2}$ state of Ba$^{+}$ ion interacting with a rotating electric
field gradient \cite{Ddm13}, all the discussion made here holds true
for any spin $3/2$ system. We choose the electric field to be zero
to avoid monopolar interaction. The dipole moment of this state is
zero because of the definite parity of the states and hence we
choose the electric quadrupole moment. Electric quadrupole moment
interact with electric field gradient. To generate geometric phases,
the electric field gradient is taken to be time dependent. The time
dependence is such that the principle axes describe a conical path
about the degeneracy point as shown in Fig. 1. The spin $3/2$
interacting with the electric field gradient maintains a time
reversal symmetry and hence there are Kramer's degeneracies in the
system, i.e., $\ket{\pm1/2}$ and $\ket{\pm3/2}$ states form two
pairs of degenerate subspaces. The Hamiltonian of quadrupole
interaction has $S_{+}^2$ terms and hence couples the $\ket{\pm1/2}$
substates. Thus the $\ket{\pm1/2}$ subspace undergoes non-Abelian
evolution. The Hamiltonian however cannot couple $\ket{\pm3/2}$
states, which have a $\Delta m=3$ and thus cannot be coupled by the
quadratic terms in angular momentum operator. Hence the
$\ket{\pm3/2}$ subspace undergoes Abelian evolution. This however is
true only for the quadrupole approximation which limits the field
expansion to $S^{2}$ terms only.\\

To induce the non-Abelian to Abelian transition, we now apply a time
dependent magnetic field along with the electric field gradient. The
magnetic field lifts the degeneracy of the system and hence drives
the system away from non-Abelian behaviour. However, for a true
transition from non-Abelian to Abelian, the underlying gauge is
required to be the same in the presence of the magnetic field. This
is achieved by making the magnetic field rotate along with the
electric field gradient. This preserves the 'geometry' of the system
even in the presence of the magnetic field as the quantization axis
remains unchanged. 'Geometry' in this context implies the
transformation which connects the diagonal basis of the Hamiltonian
with the stationary basis. Because of this constancy of 'geometry',
the underlying gauge of the evolution remains invariant. In the
following sections a detailed theoretical description of the
system's evolution under the influence of the
time varying field will be derived.\\

\begin{figure}
\includegraphics[scale=.4]{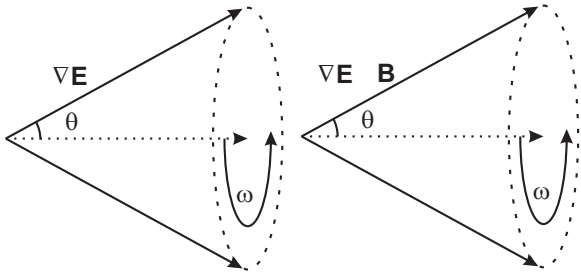}\
\label{field_fig} \caption{Configuration of the applied electric
field gradient and magnetic field for inducing non-Abelian to
Abelian transition. In the adiabatic regime, for the $\ket{\pm 1/2}$
substates, absence of magnetic field is non-Abelian while non-zero
magnetic field gives rise to Abelian geometry.}
\end{figure}


{\bf Non-Abelian Regime:} The quadrupole moment of the spin $3/2$
system interacting with an electric field gradient gives rise to the
non-Abelian regime for the $\ket{\pm1/2}$ subspace. The Hamiltonian
for a quadrupole moment- electric field gradient interaction is
given as

\begin{equation}
H_{Q}=\frac{1}{6} Q_{ij}\frac{\partial E_{i}}{\partial x_{j}},
\end{equation}
where $Q_{ij}$ is the $ij$th component of the quadrupole moment and
is defined for spin systems as $Q_{ij}=c (\frac{1}{2}(S_i S_j + S_j
S_i) -\frac{1}{3} \overrightarrow{S}^2\delta_{ij})$. $\frac{\partial
E_{i}}{\partial x_{j}}$ is the $ij$the component of the electric
field gradient tensor.\\

In our case, because of a suitable choice of the principle axes,
only the $\frac{\partial E_{z}}{\partial z}$ component of the
electric field gradient tensor contributes to the Hamiltonian of the
system. Thus, in the non-Abelian scenario, we obtain an effective
Hamiltonian given by,

\begin{equation}
H_{NA}=c(S_{z^{'}}^2-\frac{1}{3}S^{2}),
\end{equation}
where
$S_{z^{'}}=S_{x}\sin{\theta}\cos{\phi}+S_{y}\sin{\theta}\sin{\phi}+S_{z}\cos{\theta}$,
$c$ being the strength of interaction and $\phi=\omega t$, $\omega$
being the rotational frequency of the electric field gradient.\\

The eigenstates of this Hamiltonian are doubly degenerate. The two
doubly degenerate subspaces consists of $\ket{\pm\frac{3}{2}}$
corresponding to eigenvalue $c$ and $\ket{\pm\frac{1}{2}}$
corresponding to eigenvalue $-c$. Now to obtain the gauge matrices
corresponding to these sets of states through the relation
$\gamma_{mn}=i\bra{\psi_{m}}\nabla_{r}\ket{\psi_{n}}$, the
wavefunctions in the stationary frame are required. However, for the
ease of calculation, we use the Wigner D matrices to obtain the
wavefunctions in the stationary basis from the rotating basis, which
is also the diagonal basis for the Hamiltonian. The Wigner D
matrices are the transformations which connect these two bases. The
respective gauge matrices are-

\[
\gamma_{\pm 3/2}=
\begin{pmatrix}
\frac{3}{2} \cos{\theta} & 0\\
0 & -\frac{3}{2} \cos{\theta}
\end{pmatrix}
\]
and
\[
\gamma_{\pm 1/2}=
\begin{pmatrix}
\frac{1}{2} \cos{\theta} & \sin{\theta}\\
\sin{\theta} & -\frac{1}{2} \cos{\theta}\\
\end{pmatrix}
\].

From the matrices, we can see that for the subspace
$\ket{\pm\frac{1}{2}}$, we have non-zero off-diagonal elements.
Hence the degenerate subspace $\ket{\pm\frac{1}{2}}$ follows
non-Abelian evolution. In the non-Abelian regime, the eigengauge is
given by the eigenvalues of the gauge matrix,
$\pm\frac{1}{2}\sqrt{4-3\cos^{2}{\theta}}$.


{\bf Abelian Regime:} To transfer the system from non-Abelian regime
to Abelian regime, we apply a degeneracy lifting magnetic field.
However, to make any comparison between the two situations, we
require the 'geometry' of the system to remain invariant in the
presence of the magnetic field. More precisely, the connection
between the diagonal basis and the stationary frame, which is the
Wigner D matrices, should remain the same in the presence or absence
of magnetic field.

The effective Hamiltonian in the Abelian regime is given by
\begin{equation}
H_{A}=c(S_{z^{'}}^2-\frac{1}{3}S^{2})-b S_{z^{'}}.
\end{equation}

The eigenvalue of the $\ket{\pm\frac{1}{2}}$ subspace now becomes
$-c \mp \frac{1}{2}b$. However, the gauge matrix does not change, as
the geometry is kept invariant.\\

In the Abelian configuration, the adiabatic theorem leads us to
conclude that the off diagonal elements of the gauge matrix
corresponding to these set of states, do not contribute in the
physical manifestation of the phase. In this regime, the underlying
gauge is simply $\pm\frac{1}{2}\cos{\theta}$ corresponding to the
two states.

{\bf Non-Abelian to Abelian Transition:} In the two regimes of
evolution, the states, $\ket{\pm1/2}$, are governed by two different
underlying gauges given by
$\pm\frac{1}{2}\sqrt{4-3\cos{\theta}^{2}}$ for non-Abelian and
$\pm\frac{1}{2}\cos{\theta}$ for Abelian. The physical manifestation
of the gauges is obtained through phase dependent energy shifts
given by $A_{n}\omega$. Thus the variation of the energy level
shifts, on top of the energy eigenvalue, while going from
non-Abelian to Abelian is the primary signature of such a
transition.

In the true adiabatic regime ($\omega \rightarrow 0$), even the
smallest value of $b$ will drive the system from non-Abelian to
Abelian. However, for finite values of $\omega$, the system is
governed by two timescales, one depends on $\omega$, the rotational
frequency of the fields and the other depends on $b$, which
determines the splitting between the $\ket{\pm1/2}$ states. For
$\omega$ finite, Abelian regime can only be achieved for
$b\gg\omega$.

\section{Dressed State Calculations}

Unlike the systems studied so far, the system we constructed above
allows us to move continuously between the non-Abelian and Abelian
regimes. In the previous section we saw that by moving the system
from non-Abelian to Abelian regime, the off-diagonal elements loses
their physical significance. Continuous tunability of our system
allows us to investigate the dynamics of the off diagonal elements
with respect to symmetry breaking field,
in this case the magnetic field.\\

For studying the dynamics of the off diagonal elements, we begin at
the basic equation Eq. (\ref{basic_eq}) governing the evolution of
the two states. Here we work with the $\ket{\pm1/2}$ subspace. We
assume that $\omega$ is small enough compared to $c$ so that we can
neglect the coupling of the $\ket{\pm1/2}$ subspace with
$\ket{\pm3/2}$ subspace. By plugging in the values of the variables,
the governing equation for the $\ket{\pm1/2}$ substates is obtained
as

\begin{equation}
\begin{pmatrix}
\dot{C_{1}}\\
\dot{C_{2}}
\end{pmatrix}
=i
\begin{pmatrix}
\frac{\omega}{2}\cos{\theta} & \omega\sin{\theta}~e^{ibt}\\
\omega\sin{\theta}~e^{-ibt} & -\frac{\omega}{2}\cos{\theta}
\end{pmatrix}
\begin{pmatrix}
C_{1}\\
C_{2}
\end{pmatrix}.
\label{a_eq}
\end{equation}

For $b=0$, this equation governs the behaviour in the non-Abelian
regime.\\
The matrix

\begin{equation}
\mathcal{H}=
\begin{pmatrix}
\frac{\omega}{2}\cos{\theta} & \omega\sin{\theta}~e^{ibt}\\
\omega\sin{\theta}~e^{-ibt} & -\frac{\omega}{2}\cos{\theta}
\end{pmatrix}
\label{eff_ham}
\end{equation}

is like an effective Hamiltonian governing the evolution of the
states $\ket{\pm\frac{1}{2}}$ for a given value of $\omega$ and
$b$.\\


{\bf Application of Dressed State Method:} The dressed state
approach, which is a derivative of the Floquet Theorem of
differential equations with periodic co-efficients, takes into
account the full time dependence and allows us to obtain the true
eigenvalues, considering all the effects of the time dependence.
Even though it was first developed to deal with atom photon
interaction, it can be generalized to any equation with periodic
co-efficients. For a detailed description of the mathematical
algorithm applied here to
obtain the eigenvalues of Eq. (\ref{eff_ham}), please refer to \cite{Meyer09}.\\

The eigenvalue obtained using the dressed state algorithm provides
the complete picture including the effect of phase dependent energy
shifts of the level. It also allows us to obtain the complete
dependence of the geometric phase on $b$ and $\omega$ and thus
letting us probe not only the non-Abelian and Abelian
limit but the entire behaviour of the system.\\

\begin{figure*}[t]
\subfigure[]{\includegraphics[scale=1.7]{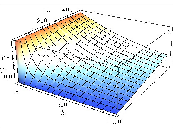}}
\subfigure[]{\includegraphics[scale=.4]{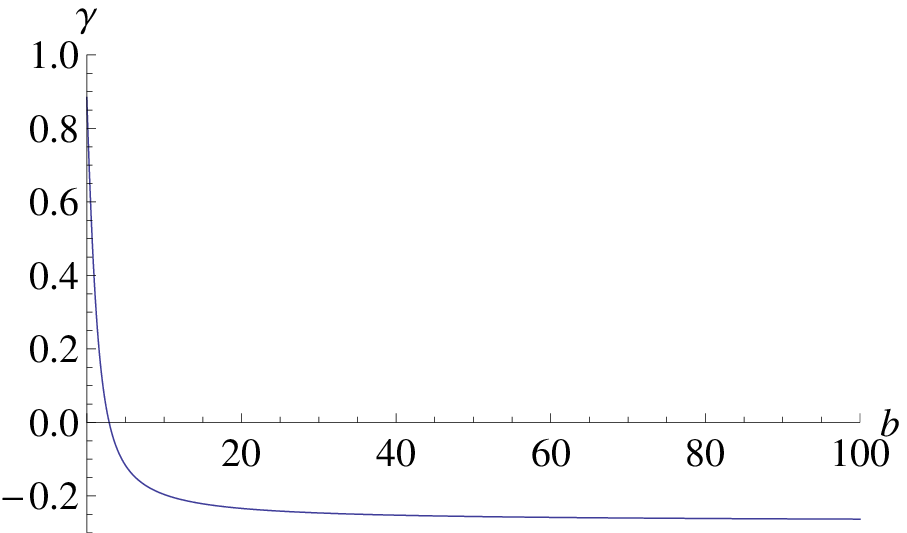}}
\subfigure[]{\includegraphics[scale=.4]{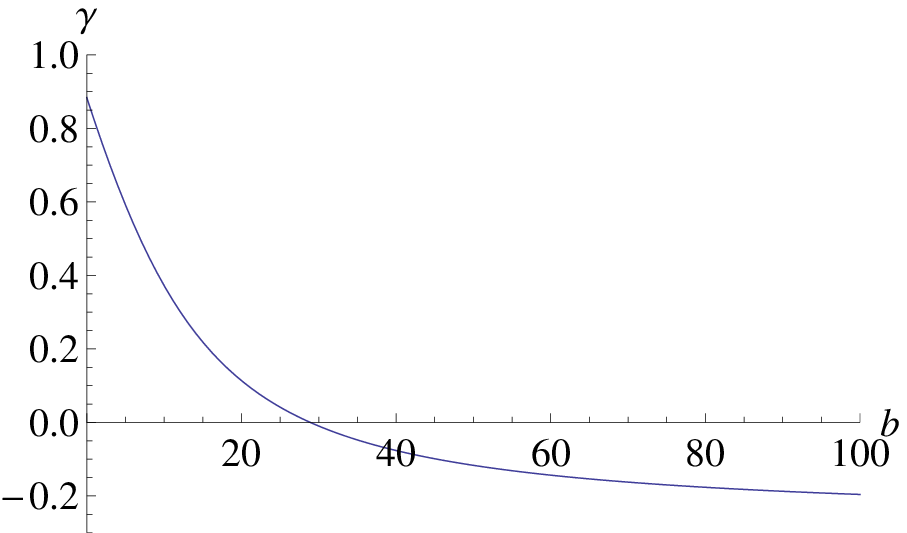}}
\subfigure[]{\includegraphics[scale=.4]{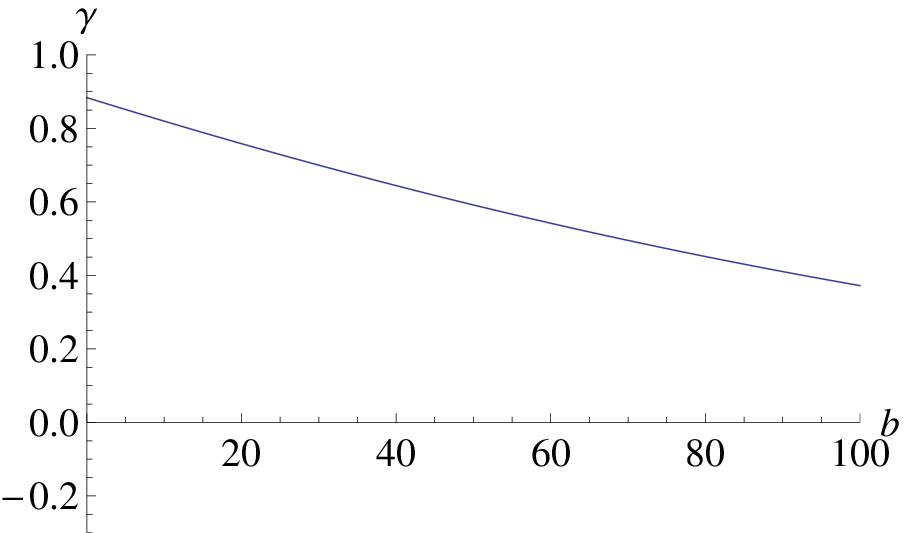}}
 \label{delwn}
\caption{Figure demonstrates the non-Abelian to Abelian transition.
Figure (a) depicts the three dimensional dependence of the evolution
gauge on $b$ and $\omega$. Figures (b),(c) and (d) show the
behaviour of the system moving away from the non-Abelian point for
angular frequencies of $1$, $10$ and $100$ Hz. As can be seen, with
increase of rotational frequency, the transition of the system is
much more slow. The non-abelian behaviour is 'retained' for higher
values of magnetic field for a higher angular frequency. The figure
also demonstrates that by selecting appropriate values of $b$ and
$\omega$, we can bring control the behaviour of the system
precisely, with the phase acquired defined only by the pair of
values of $b$ and $\omega$. For these graphs, the value of $\theta$
is kept as $57.3^{o}$. Magnetic field $b$ and angular frequencies
$\omega$ are expressed in $Hz$.}
\end{figure*}


To apply the dressed state method, we assume an ansatz of the form

\begin{equation}
\begin{pmatrix}
\dot{C_{1}}\\
\dot{C_{2}}
\end{pmatrix}
=
\begin{pmatrix}
\alpha_{1}(t) e^{-i\omega_{+}t}\\
\alpha_{2}(t) e^{-i\omega_{-}t}
\end{pmatrix}
.
\label{alp_eq}
\end{equation}

Now by inserting Eq. \ref{alp_eq} into Eq. \ref{a_eq}, we obtain the
following equation for $\alpha_{1}$ and $\alpha_{2}$

\begin{widetext}
\begin{equation}
i
\begin{pmatrix}
\dot{\alpha_{1}}\\
\dot{\alpha_{2}}
\end{pmatrix}
=
\begin{pmatrix}
-\frac{\omega}{2}\cos{\theta}-\omega_{+} & -\omega\sin{\theta}~e^{i(b+\omega_{+}-\omega_{-})t}\\
-\omega\sin{\theta}~e^{i(b+\omega_{+}-\omega_{-})t} &
\frac{\omega}{2}\cos{\theta}-\omega_{-}
\end{pmatrix}
\begin{pmatrix}
\alpha_{1}\\
\alpha_{2}
\end{pmatrix}.
\label{aa_eq}
\end{equation}
\end{widetext}

Now if we choose $\omega_{\pm}=\mp\frac{b}{2}$, then the above
$2\times2$ matrix becomes time independent and all the information
about the time dependence of the system becomes encoded in behaviour
of $\alpha_{1}(t)$ and $\alpha_{2}(t)$. Such a choice of the values
of $\omega_{\pm}$ converts the above equation into a time
independent problem, with an effective Hamiltonian given by

\begin{equation}
\mathcal{H}_{D}=
\begin{pmatrix}
-\frac{\omega}{2}\cos{\theta}+\frac{b}{2} & -\omega\sin{\theta}\\
-\omega\sin{\theta} & \frac{\omega}{2}\cos{\theta}-\frac{b}{2}
\end{pmatrix}.
\label{dress_ham}
\end{equation}

The Hamiltonian in Eq. (\ref{dress_ham}) is the dressed form of the
effective Hamiltonian given by Eq. (\ref{eff_ham}). The advantage is
that we converted the time dependent problem into an effective time
independent problem thus allowing us to capture the complete
behaviour of the system through the
eigenvalues of the dressed Hamiltonian.\\
The eigenvalues of $\mathcal{H}_{D}$ are given by $\pm
\frac{\omega}{2}\sqrt{4
\sin{\theta}^2+\cos{\theta}^{2}+(\frac{b}{\omega})^{2}-2
\cos{\theta}\frac{b}{\omega}}$. The complete solution for $C_{1}$
and $C_{2}$ is given by

\begin{equation}
\begin{pmatrix}
{C_{1}}\\
{C_{2}}
\end{pmatrix}
=
\begin{pmatrix}
C_{1}(0) e^{-i\lambda t}\\
C_{2}(0) e^{+i\lambda t}
\end{pmatrix},
\label{alp_eq_1}
\end{equation}

where,

\begin{equation}
\lambda=\frac{\omega}{2}(\sqrt{4
\sin{\theta}^2+\cos{\theta}^{2}+(\frac{b}{\omega})^{2}-2
\cos{\theta}\frac{b}{\omega}}-\frac{b}{\omega}).
\end{equation}

Here $\lambda$ represents the phase dependent energy shift of the
levels. As can be seen, this shift is of the form
$\gamma_{n}\omega$. In the pure non-Abelian or Abelian regime, the
value of $\gamma_{n}$ is independent of $b$ or $\omega$. However, in
intermediate region, this gauge of the system depends on both the
value of the degeneracy lifting field
as well as the frequency of evolution.\\

To obtain the non-Abelian limit, we put $b=0$ and obtain the
familiar non-Abelian gauge eigenvalues given by
\begin{equation*}
\pm \sqrt{4 - 3 \cos^2{\theta}}.
\end{equation*}

We can reach the Abelian limit by putting $\omega\rightarrow0$ for
any value of $b\neq0$. However, physically, the exact value of
$\omega$ required to reach the Abelian limit, depends on the value
of $b$ and hence a more suitable limit for the Abelian regime is
$\frac{b}{\omega}\gg1$. In this limit, the gauge tends to
\begin{equation}
\pm\frac{1}{2}\cos{\theta}.
\end{equation}

It should be mentioned here that the choice of
$\omega_{\pm}=\pm\frac{b}{2}$ is also a valid choice of the ansatz.
However, this choice represents a the opposite sense of rotation of
the fields. In principle, these two choices physically corresponds
to a difference of $\pi$ of the angle between $\omega$ and $b$.



\section{Perturbative analysis of different contributions}

In the previous section, we have obtained the complete behavior of
the system. However, the dressed state approach did not reveal the
underlying physical phenomenon governing the transition region. The
physical processes controlling the two limits is however known. Now
to obtain the physics of the transition region, we approach
perturbatively from the two extremes and try to figure out the
physical processes driving the system away from the two limits.\\

In this section our goal is to capture the response of the system
due to small changes in system parameter($b$ or $\omega$) from its
two extreme limits. We apply perturbation theory to achieve this.
The key point is to choose unperturbed Hamiltonian in the two
regimes. As the non-Abelian and Abelian regimes are very different
in nature one should not expect to use the same unperturbed
Hamiltonian to describe both. We work with the dressed state
Hamiltonian, where the problem is reduced to a time independent
situation. We use this 'dressed' Hamiltonian, to identify the
unperturbed Hamiltonians governing the behaviour of the system in
the two limits. Other than the unperturbed Hamiltonian, whatever is
left, is treated as the perturbation.\\

{\bf Perturbation in the Abelian Limit:} The Abelian limit
corresponds to the situation where $b\gg\omega$. In this condition,
the phase dependent energy shift is given by
$\frac{\omega}{2}\cos{\theta}$. The effective time independent
Hamiltonian which can describe this system, including the phase
dependent energy shifts, can be written as

\begin{equation}
H_{D}^{A}=
\begin{pmatrix}
\frac{1}{2}\cos{\theta}\omega -\frac{b}{2} & 0\\
0 & -\frac{1}{2}\cos{\theta}\omega +\frac{b}{2}
\end{pmatrix}
\label{h_dress_o}
\end{equation}

Now to study the deviation of the system from the Abelian limit, we
rewrite the total dressed Hamiltonian as

\begin{equation*}
\mathcal{H}_{D}={H}_{D}^{A}+\delta H_{D}^{A},
\end{equation*}

where $\delta H_{D}^{A}=\mathcal{H}_{D}-{H}_{D}^{A}$ is the
perturbing Hamiltonian. The form of $\delta H_{D}$ comes out as

\begin{equation}
\delta{H}_{D}^{A}=
\begin{pmatrix}
0 & \omega \sin{\theta}\\
\omega \sin{\theta} & 0
\end{pmatrix}.
\label{h_dress_p}
\end{equation}

Now we calculate the terms of the perturbation series using this
$\delta H_{D}^{A}$. The first order contribution of this
perturbation being zero, the leading term of the perturbation series
is the second order contribution which is
\begin{equation}
E_{D}^{''}=-\frac{\sin^{2}{\theta}}{\cos{\theta}-\frac{b}{\omega}}.
\end{equation}

We can thus have a handle on the underlying physical processes
governing the level shift of the system from the Abelian behaviour.
As the perturbation series reveals, the 'non-Abelian' perturbation
does not effect the energy levels of the unperturbed states.
However, it causes a population transfer between the eigenstates as
is given by a non-zero second order term. We also notice that
decreasing the value of $\frac{b}{\omega}$, which we know takes the
system away from Abelian behaviour, also increases the coupling
between the two
states.\\

This perturbation however fails when $\frac{b}{\omega}$ approaches
$\cos{\theta}$. This is because for $b=\omega\cos{\theta}$, the
second order contribution has a singularity. This point indicates a
deviation of the guiding physics from the Abelian behaviour. For
$b>\omega|\cos{\theta}|$, the Hamiltonian decomposition used above
holds true.\\

\begin{figure}[t]
\centering
\includegraphics[scale=.8]{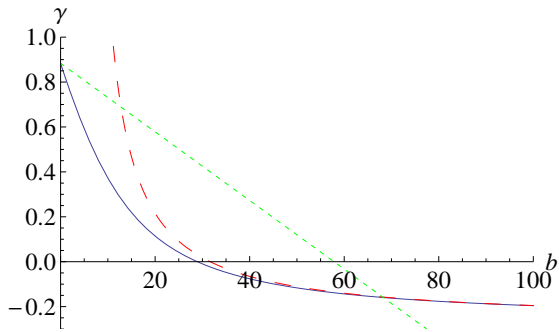}
\label{device} \caption{The figure depicts the results of the
perturbative analysis with respect to the exact behaviour. The
dashed line corresponds to the perturbation in the abelian regime.
As can be seen, the perturbation mimics the exact behaviour till it
reaches close the $b=\omega\cos{\theta}$ limit of the perturbation.
The dotted line denotes the perturbation in the non-Abelian regime.
As can be seen, the second order perturbation in this regime does
not have a strong correspondence with the exact behaviour. This
hints towards significant contribution from higher order
perturbation terms, which is beyond the scope of this work. For this
graph $\theta=57.3^{0}$ and $\omega=10$ Hz.}
\end{figure}

{\bf Perturbation in the Non-Abelian Limit:} The non-Abelian limit
corresponds to $b\ll\omega$. At the non-Abelian point, that is
$b=0$, the phase dependent energy shift is given by
$\frac{\omega}{2}\sqrt{4-3\cos^2{\theta}}$. As in the Abelian limit,
in the non-Abelian limit, the Hamiltonian is given by

\begin{equation}
H_{D}^{NA}=
\begin{pmatrix}
\frac{1}{2}\cos{\theta}\omega  & \sin{\theta}\omega\\
\sin{\theta}\omega & -\frac{1}{2}\cos{\theta}\omega
\end{pmatrix}
\label{h_dress_o}
\end{equation}

We again write the total dressed Hamiltonian as

\begin{equation*}
\mathcal{H}_{D}={H}_{D}^{NA}+\delta H_{D}^{NA}
\end{equation*}

where $\delta H_{D}^{NA}=\mathcal{H}_{D}-{H}_{D}^{NA}$ is the
perturbing Hamiltonian in the Non-Abelian limit. $\delta H_{D}^{NA}$
is given as

\begin{equation}
\delta H_{D}^{NA}=
\begin{pmatrix}
-\frac{b}{2} & 0\\
0 & +\frac{b}{2}
\end{pmatrix}
.
\label{h_dress}
\end{equation}

Now we calculate the perturbating terms using this $\delta
H_{D}^{NA}$. We find now that unlike the Abelian limit, in this
case, the first order perturbation is non-zero whereas the second
order contribution is zero. The first order contribution is given as

\begin{equation}
E_{D}^{'}=\frac{b}{\omega}\frac{\cos{\theta}}{2\sqrt{4-3\cos^2{\theta}}}.
\end{equation}

This series illuminate the fact that in the deviation of the system
from non-Abelian behaviour, level shift plays the major role. The
increasing energy gap between the levels however leads to a reduced
rate of population transfer and thus drives the system away from the
non-Abelian behaviour.
\section{Sensitivity of Geometric Phase to Parameter Fluctuations}

Having explored both the regions perturbatively, we can now analyze
the influence of fluctuations of different external parameters on
the geometric phase.\\

To develop a general framework, let us assume that $b^{'}= b+\delta
b$ and $\omega^{'}=\omega+ \delta \omega$, where $\delta b$ and
$\delta \omega$ are the fluctuations of $b$ and $\omega$
respectively. The phase at the two extremes, that is the Abelian and
non Abelian limits are independent of the values of $b$ and
$\omega$. However in the intermediate regions, it is dependent on
these parameters. In general, we can write the phase as follows

\begin{equation}
\gamma(b,\omega)=\gamma_{0}+\gamma^{'}(b,\omega)
\end{equation}

where $\gamma_{0}$ can be either the Abelian or non-Abelian phase
and $\gamma_{'}$ is the perturbative deviation in each limit.
$\gamma_{0}$ is independent of $\omega$ and $b$ in both the
limits.\\

Now to study the effect of fluctuation of each parameter on the
geometric phase, we apply the above mentioned substitution and
obtain

\begin{equation}
\gamma(b^{'},\omega^{'})=\gamma_{0}+\gamma^{'}(b,\omega)+\frac{\partial
\gamma^{'}}{\partial \omega}\delta \omega + \frac{\partial
\gamma^{'}}{\partial b}\delta b
\end{equation}

The effects of the fluctuations of these parameters on the geometric
phase are obtained through $\frac{\partial \gamma^{'}}{\partial
\omega}$ and
$\frac{\partial \gamma^{'}}{\partial b}$.\\

In the Abelian limit, we have

\begin{equation}
\frac{\partial \gamma ^{'}}{\partial
\omega}=\frac{\sin^{2}{\theta}}{b}
\end{equation}

and

\begin{equation}
\frac{\partial \gamma ^{'}}{\partial
b}=\frac{\omega\sin^{2}{\theta}}{b^{2}}.\\
\end{equation}
In the above equations it is assumed $\frac{b}{\omega}\gg 1$ which
holds true in the Abelian regime.\\
For Non-Abelian limit,
\begin{equation}
\frac{\partial \gamma ^{'}}{\partial
\omega}=\frac{b\cos{\theta}}{2\omega^{2}\sqrt{4-3\cos^{2}{\theta}}}\\
\end{equation}
and
\begin{equation}
\frac{\partial \gamma ^{'}}{\partial
b}=\frac{\cos{\theta}}{2\omega\sqrt{4-3\cos^{2}{\theta}}}.
\end{equation}

From the above four equations, depicting the effect of fluctuation
of parameters on the geometric phase, it is evident that the role of
$\omega$ and $b$ are interchanged in the non-Abelian and the Abelian
limits. While the non-Abelian limit is more sensitive to magnetic
field fluctuations, the Abelian limit on the other hand is more
sensitive to fluctuations of the angular frequency.\\

Experimentally usually magnetic field noise is one of the biggest
source of dephasing for quantum systems. From that point of view, we
can say that the phase fluctuations due to magnetic field noise will
be much higher in the non-Abelian regime than compared to the
Abelian regime and thus the Abelian limit is much more robust
against fluctuation of magnetic field. However, if in a certain
situation, robustness against rotational frequency is required, then
non-Abelian limit is a much better choice than
Abelian limit.\\

\section{Discussion and Conclusion}

In this work we have performed an extensive study of a system as it
is continuously moved from Abelian regime to Non-Abelian regime.
Although it was known that Abelian regime is signified by non
transfer of population and non-Abelian system by population
transfer, we have for the first time shown the dynamics of the
system with respect to a symmetry breaking field, driving the system
from one regime to another.\\
The dressed state approach revealed the exact dynamics of the system
and at the same time allowed us to probe the underlying mechanisms a
play using the method of perturbation.\\
The perturbative approach also helped us to gain insight into
robustness of geometric phase to external parameter fluctuations. As
can be seen the non-Abelian limit is more susceptible to magnetic
field fluctuations whereas Abelian limit is prone to fluctuations
arising from angular frequency fluctuations. Thus a detailed
understanding of Abelian and non-Abelian evolutions and the
behaviour off the system in between can in turn lead to better
designing of architecture for implementation of quantum computation
protocols. The two field system provides a greater handle on phase
engineering requirements for quantum technology purposes.\\

\section{Acknowledgements}

The authors would like to thank Mr. Sanjib Ghosh for extensive
discussion both on the scientific as well as presentation aspect of
this paper. We would also like to thank CQT for financial support.

\end{document}